\documentclass[aps,prl,floatfix,footinbib,twocolumn,superscriptaddress]{revtex4-1}
\usepackage{bm}
\usepackage{graphicx}
\usepackage{epstopdf}
\usepackage{xcolor}
\usepackage{color}
\usepackage{lineno}
\usepackage{amssymb,amsfonts,amsmath}
\DeclareFontFamily{U}{euc}{}
\DeclareFontShape{U}{euc}{m}{n}{<-6>eurm5<6-8>eurm7<8->eurm10}{}%
\DeclareSymbolFont{AMSc}{U}{euc}{m}{n} 
\DeclareMathSymbol{\umu}{\mathord}{AMSc}{"16}
\usepackage[printwatermark]{xwatermark}
\usepackage[bookmarks=false, pdfstartview={FitH}, colorlinks = true, citecolor = blue, linkcolor = blue, urlcolor = blue,hyperfootnotes=true]{hyperref}
\usepackage{cleveref}

\makeatletter
\renewcommand{\@fnsymbol}[1]{\ensuremath{\ifcase#1\or \dagger\or *\or \ddagger\or
   \mathsection\or \mathparagraph\or \|\or **\or \dagger\dagger
   \or \ddagger\ddagger \else\@ctrerr\fi}}
\makeatother

\begin{document}

\def\PENN{\small Department of Mechanical Engineering \& Applied Mechanics, University of Pennsylvania, Philadelphia, PA 19104, USA}
\def\GTown{\small Department of Physics, and Institute for Soft Matter Synthesis \& Metrology, Georgetown University, 3700 O Street NW, Washington, D.C. 20057, USA}
\def\Edin{\small SUPA, School of Physics and Astronomy, The University of Edinburgh, James Clerk Maxwell Building, Peter Guthrie Tait Road, Edinburgh, EH9 3FD, United Kingdom}

\newcommand{\degreesC}{\,^{\circ}{\rm C}}
\newcommand{\degrees}{\,^{\circ}}

\title{Polymers in Swarming Bacterial Turbulence} 
\author{Ranjiangshang Ran}
\thanks{R.R. and D.A.G contributed equally to this work.}
\affiliation{\PENN}
\author{David A. Gagnon}
\thanks{R.R. and D.A.G contributed equally to this work.}
\affiliation{\PENN}
\affiliation{\GTown}
\author{Alexander Morozov}
\affiliation{\Edin}
\author{Paulo E. Arratia}
\thanks{parratia@seas.upenn.edu\vspace{+0.6ex}}
\affiliation{\PENN}

\date{\today}

\begin{abstract}
We experimentally investigate the effects of polymer additives on the collective dynamics of swarming \textit{Serratia marcescens} in quasi two-dimensional (2D) liquid films. We find that even minute amounts of polymers ($\leq$ 20 ppm) can significantly enhance swimming speed and promote large-scale coherent structures. Velocity statistics show that polymers suppress large velocity fluctuation, transforming the velocity distributions from super-Gaussian to Gaussian. Spatial and temporal correlation functions suggest that polymers increase both the size and lifetime of flow structures. The energy spectra show an exponential decay at low wavenumbers, with a characteristic length scale increasing with polymer concentration. Overall, these result show polymers can mediate bacteria interaction and promote large-scale coherence in dense active suspensions.

\end{abstract}


\maketitle 


Even simple life forms, like bacteria and protozoa, can exhibit complex behaviors such as swarming \cite{Kearns_NRM_2010,Ariel_NC_2015}, quorum sensing \cite{Waters_QuorumSensing_2005,Daniels_Quorum_PNAS_2006}, and biofilm formation \cite{Costerton_Science_1999,Hall_Stoodley_NRM_2004,desai_biofilms_2020}. At sufficiently high cell densities, microorganisms can communicate chemically \cite{Berg_Nature_1995,Kalinin_BioJ_2009} and hydrodynamically \cite{Sokolov_PRL_2007,Sokolov_PRL_2012}, and move together in a coordinated manner known as collective motion \cite{Koch_ARFM_2011}. An intriguing phenomenon is the emergence of turbulent-like features in bacterial suspensions that include large-scale coherence \cite{Dombrowski_PRL_2004,Sokolov_PRL_2007,Gachelin_NJP_2014}, strongly fluctuating velocity \cite{Guasto_PRL_2010, Rushkin_PRL_2010}, and anomalous diffusivity \cite{Ariel_NC_2015,Mukherjee_PRL_2021}. Due to the qualitative similarity to turbulence at high Reynolds numbers, these behaviors are often referred as ``bacterial turbulence'' \cite{Dombrowski_PRL_2004,Sokolov_PRL_2007,Koch_ARFM_2011,Sokolov_PRL_2012,Sokolov_PRE_2009,Cisneros_PRE_2011,Wensink_PNAS_2012,Dunkel_PRL_2013,Gachelin_NJP_2014}. Both theoretical and numerical studies based on continuum theory \cite{Dunkel_PRL_2013,Wensink_PNAS_2012,Saintillan_PRL_2008,Subramanian_JFM_2009,Hohenegger_PRE_2010,Saintillan_CRP_2013,Saintillan_SM_2017,Saintillan_JCP_2019} as well as discrete swimming particles \cite{Graham_PRL_2005,Saintillan_PRL_2007,Underhill_PRL_2008,Wolgemuth_BioJ_2008,Graham_JPCM_2009,Leoni_PRL_2010,Lushi_PNAS_2014,Krishnamurthy_JFM_2015,Morozov_PRL_2017,Morozov_SM_2019} have shown that hydrodynamic effects alone can capture some of the main features of bacterial turbulence, even without biochemical interactions.

Microorganisms often live in fluid environments where (bio)polymers are present \cite{Lauga_PoF_2007}. For instance, bacteria can secrete slime to reduce friction while swarming across a solid surface \cite{Angelini_PNAS_2009}, and produce protective exopolymeric matrix during the formation of biofilms \cite{Seminara_PNAS_2012}. How the presence of polymer molecules in the fluid medium affect the swimming behavior of single microorganisms has received much attention in the past decade or so; both enhancement \cite{Teran_PRL_2010,Morozov_Poon_PNAS_2014, Patteson2015,Yeomans_NatPhys_2019} and hindrance \cite{Shen_PRL_2011,Liu_PNAS_2011,Becca_PRL_2014, Qin2015} in swimming speed have been found depending on the often nonlinear interaction between the swimmer kinematics, velocity fields, and fluid rheological properties. Less explored, however, are the effects of polymers on the collective behavior of swimming microorganisms. Recently, large spatial-temporal ordered structures (vortices) were found in bacterial suspensions inside droplets containing viscoelastic fluids (DNA suspensions)\cite{Liu_Nature_2021}; these ordered structures disappear as polymer concentration is decreased. In the dilute regime, numerical simulations on the collective dynamics of rod-like swimmers, on the other hand, show that fluid elasticity suppress velocity fluctuations and break down large-scale flow structures \cite{Li_PRL_2016}, while simulations based on mean field theory suggest that fluid elasticity can mediate hydrodynamic interactions and lead to larger coordinated structures \cite{underhill_PRE_2011}. These expectations have yet to be tested in experiments. As a result, it is still unclear how the presence of polymer molecules affect the collective behavior of swimming microorganisms.

\begin{figure}[b!]\label{fig1}
\centering
\includegraphics[width=3.37in]{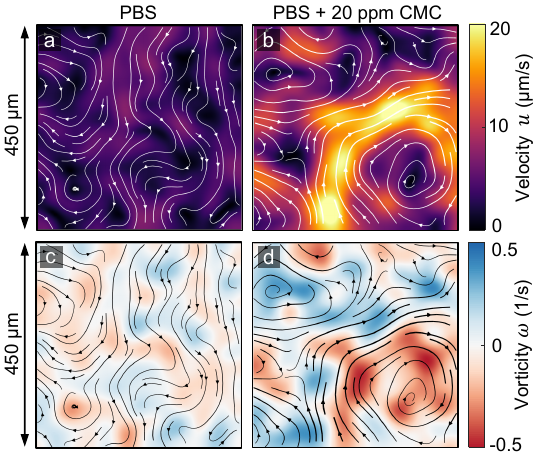}
\vspace{-2ex}
\caption{PIV flow fields of swarming \textit{S. marcescens} in the buffer and polymeric solutions. (a, b) Velocity magnitude fields $u$ and (c, d) vorticity fields $\omega$, for the PBS buffer (left) and polymeric solutions with 20 ppm of CMC (right). Solid lines are streamlines computed from instantaneous velocity fields. See Supplemental Material \cite{Supp_Mat} for movies.}
\end{figure}

In this study, we experimentally investigate the effects of polymer additives on the collective dynamics of bacterial suspensions in quasi-2D liquid films. Results show that minute amounts of polymers ($\le20$ ppm) can significantly enhance bacterial propulsion speed and promote large-scale coherent structures. Velocity statistics show that polymers additives suppress velocity fluctuations and change the velocity distribution from super-Gaussian to Gaussian. Spatial and temporal correlation functions suggest that the size and lifetime of the flow structures increase by up to 50\% in the presence of polymers. Energy spectra show an exponential decay with a characteristic length scale that increases with polymer concentration. Overall, our results provide insights into bacterial collective motion in complex fluid environments where polymers are present.

Experiments are performed on a strain (ATCC 274) of \textit{Serratia marcescens}, a rod-shaped bacterium that is on average 2~$\umu$m long and 0.8~$\umu$m in diameter. When cultivated on agar plates \cite{Supp_Mat}, the bacteria differentiate into swarmer cells with additional (10 to 100) flagella and elongated bodies of $\ge5~\umu$m \cite{Harshey_JBac_1990}. The swarmer \textit{S. marcescens} can move at approximately $30~\umu$m/s on agar plates, and exhibit super-diffusive trajectories through L\'evy walks \cite{Ariel_NC_2015}. These swarmer cells are then transferred into either a Newtonian buffer (phosphate-buffered saline, PBS) or a polymeric solution. Polymeric solutions are prepared by diluting a carboxymethyl cellulose (CMC, $7\times10^5$ Da) stock solution in the PBS buffer to final concentrations of 5, 10, and 20 ppm. Note that the highest polymer concentration is well below the overlap concentration $c^*$ ($\le0.2\%\,c^*$). A 2-$\umu$L drop of PBS or CMC solution containing swarmer \textit{S. marcescens} of a volume fraction of roughly $30\,\textrm{--}\,40$\% is placed in a thin-film apparatus \cite{Patteson2015, Qin2015, Supp_Mat}, and stretched into a 1-cm${}^2$ large and 40-$\umu$m thin film. Images are taken using bright-field microscopy and a CMOS camera (Flare 4M180) at 24~frame/s. Velocity fields are obtained using particle imaging velocimetry (PIV, see \cite{Gagnon_PRL_2020}), with a total number of 6400 or $80\times80$ interrogation windows, each of a size of $25\times25$~pixel or $7.0\times7.0$~$\umu$m${}^2$.

Figure \hyperref[fig1]{1} shows experimental velocity and vorticity fields for the buffer (PBS) and the 20 ppm CMC solution (see movies in \cite{Supp_Mat}); instantaneous streamlines are overlaid on fields to better visualize structures. The flow fields show that the addition of polymers significantly increases the overall speed $u$ of the swarming bacterial suspension; the maximum suspension speed is nearly doubled from 10~$\umu$m/s in the buffer to 20~$\umu$m/s in the polymeric solution [Fig. \hyperref[fig1]{1(a)} and \hyperref[fig1]{1(b)}]. While it has been previously found that a single microbe can swim faster in polymeric solutions \cite{Teran_PRL_2010,Morozov_Poon_PNAS_2014, Yeomans_NatPhys_2019, Patteson2015}, the flows observed here are not merely scaled up by a higher speed (of an individual bacterium). If that were the case, one would expect the flow structures to remain roughly of the same size. Here, on the other hand, we find that the flow structures length-scale increases with the addition of polymers, as shown by the vorticity fields [Fig. \hyperref[fig1]{1(c)} and \hyperref[fig1]{1(d)}]. This indicates that bacterial collective motion in these ultra-dilute polymeric fluids have distinct underlying flow structures from those in Newtonian fluids.

\begin{figure}[t!]\label{fig2}
\centering
\includegraphics[width=3.37in]{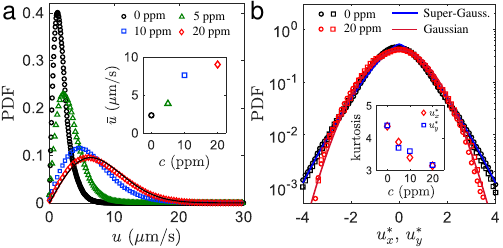}
\caption{(a) PDFs of the velocity magnitude $u$ for different polymer concentrations $c$. The black solid curve is a Rayleigh distribution fit for the 20 ppm CMC data. Inset: root mean square (RMS) velocity $\bar{u}$ versus $c$. (b) PDFs of the velocity components $u_x^*$ (circles) and $u_y^*$ (squares), normalized to have a zero-mean and a unity-variance. Solid curves are Gaussian and super-Gaussian fits to the data. Inset: kurtosis of the PDFs of $u_x^*$ and $u_y^*$ as a function of $c$.}
\end{figure}

The effects of polymers on the flow structures of collective motion are further quantified by calculating the probability density functions (PDFs) of the velocity magnitude fields. We find that the addition of 20 ppm of CMC more than doubles the maximum swarming speed $u$ [Fig. \hyperref[fig2]{2(a)}], and roughly triples the mean speed $\bar{u}$ [Fig. \hyperref[fig2]{2(a)}, inset]. Here, $\bar{u}$ is defined by the root mean square (RMS) velocity, $\bar{u}=\langle u^2\rangle^{1/2}$, where $\langle\cdot\rangle$ denotes the spatiotemporal ensemble average. We note that these PDFs are not simply rescaled, rather, they follow different statistical distributions. As polymer is added to the suspensions, the PDFs of the velocity magnitude tend to be a Rayleigh distribution [black curve in Fig. \hyperref[fig2]{2(a)}]; such distribution arises when the two velocity components follow independent Gaussian distributions. Previously, it was reported that the speed distribution in dilute suspensions of swimming peritrichous bacteria follows the Schultz distribution \cite{Wilson_PRL_2011}. Here, instead, we find Rayleigh and non-Rayleigh speed distributions for polymeric solutions and Newtonian buffer, respectively. This result also suggests that the PDFs of velocity components are Gaussian for the 20 ppm CMC case and non-Gaussian for the buffer case.


To test this idea, we compute the PDFs of the in-plane velocity components $u_x$ and $u_y$ for the 0 ppm (PBS) and 20 ppm CMC cases [Fig. \hyperref[fig2]{2(b)}]. For better contrast, the velocity components $u_x^*$ and $u_y^*$ are normalized to have a mean of zero and a standard deviation of unity. Importantly, we find no noticeable difference between the PDFs of x- and y-velocity components, suggesting the in-plane motion of bacteria is statistically isotropic. In the buffer case (0 ppm), the velocity distributions are broadened, with heavy tails at high velocities. A generalized Gaussian function fitting, $N\exp(-c|u_{x,y}^*|^{\beta})$, reveals that the PDFs are super-Gaussian with $\beta\approx1.4$. In contrast, such tails are absent in the polymeric case (20 ppm), and the PDFs are approximately Gaussian with $\beta\approx2.0$. That is, polymer additives seem to suppress the tails of velocity distributions by decreasing velocity fluctuations.

The suppression of tails in the velocity PDFs can be characterized by the kurtosis of velocity components [Fig. \hyperref[fig2]{2(b)}, inset]. The kurtosis is 3 for Gaussian distributions, greater than 3 for super-Gaussian and less than 3 for sub-Gaussian distributions. We find that as the polymer concentration increases, the kurtosis of velocity components decreases from $\sim4.5$ in the buffer (0 ppm) to $\sim3$ in the polymeric fluid (20 ppm). The decrease in kurtosis suggests that polymers suppress the intermittency of velocity fluctuations, making the flows in 2D bacterial turbulence more uniform. The observed decrease in velocity fluctuation with polymers is consistent with previous numerical simulations \cite{Li_PRL_2016}.

\begin{figure}[b!]\label{fig3}
\centering
\includegraphics[width=3.37in]{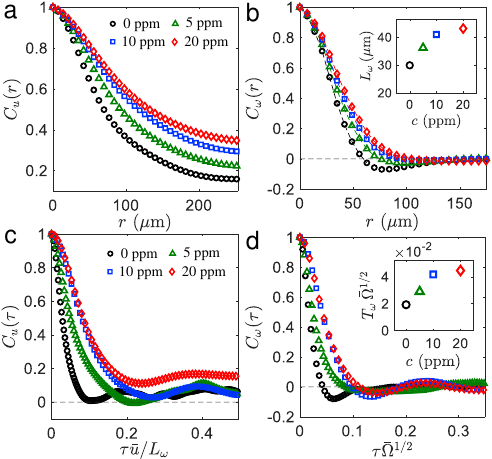}
\caption{(a, b) Spatial correlation functions of (a) velocity and (b) vorticity. Inset: vorticity integral length scale $L_\omega$. (c, d) Temporal correlation functions of (c) velocity and (d) vorticity. The time lag $\tau$ is normalized by the eddy turnover time $L_\omega/\bar{u}$ in (c), and by the enstrophy time scale $\bar{\Omega}^{-1/2}$ in (d). Inset: normalized vorticity integral time scale $T_\omega\,\bar{\Omega}^{1/2}$.}
\end{figure}

Next, we quantify the size of flow structures by the spatial correlation functions of the velocity $\mathbf{u}$ and the vorticity $\omega$, defined as: $C_u(r)=\langle\mathbf{u}(\mathbf{x})\cdot\mathbf{u}(\mathbf{x}+\mathbf{r})\rangle/\langle u^2\rangle$ and $C_\omega(r)=\langle\omega(\mathbf{x})\,\omega(\mathbf{x}+\mathbf{r})\rangle/\langle\omega^2\rangle$. We note that here $\mathbf{u}(\mathbf{x})$ denotes the local velocity of bacteria at the position $\mathbf{x}$, and comprises two contributions: self-propulsion and advection created by other bacteria. As polymers are added, we find that the velocity fields are increasingly correlated over a distance of $250~\umu$m [Fig. \hyperref[fig3]{3(a)}]. Similar increases in spatial correlations are also found in the vorticity fields [Fig. \hyperref[fig3]{3(b)}]. The average vortex size can be estimated by the integral length scale of vorticity $L_\omega$, defined by the convergent integral: $L_\omega=\int_0^\infty C_\omega(r)\,dr$. The inset of Fig. \hyperref[fig3]{3(b)} shows that the average vortex size increases with polymer concentration; $L_\omega$ increases by as much as $50\%$ (relative to the buffer case) by adding 20 ppm of polymers. This result is in contrast to previous numerical studies \cite{Underhill_JNNFM_2014,Li_PRL_2016}, where the addition of polymers seems to break down large-scale flow structures and reduce the structure size. We note that a much lower swimmer number density was used in the simulations than in the current experiments, which may be responsible for the discrepancy. 

The lifetime of flow structures are examined by the temporal correlation functions of velocity $\mathbf{u}$ and vorticity $\omega$, defined as: $C_u(\tau)=\langle\mathbf{u}(t)\cdotp\mathbf{u}(t+\tau)\rangle/\langle u^2\rangle$ and $C_\omega(\tau)=\langle\omega(t)\,\omega(t+\tau)\rangle/\langle \omega^2\rangle$. To compensate for the increases in flow speed and structure size with polymers, the time lag $\tau$ for velocity correlation is rescaled by the eddy turnover time, $L_\omega/\bar{u}$, where $\bar{u}$ is the RMS velocity [see Fig. \hyperref[fig2]{2(a)}, inset]. An increase in velocity temporal correlations is found [Fig. \hyperref[fig3]{3(c)}] up to half of an eddy turnover time ($\sim5$~s), suggesting that polymers increase the average lifetime of flow structures. The vorticity fields are also increasingly correlated in time with the addition of polymers [Fig. \hyperref[fig3]{3(d)}]. Here, the time lag $\tau$ is normalized by the enstrophy time scale, $\bar{\Omega}^{-1/2}$, where enstrophy is defined by the mean square vorticity, $\bar{\Omega}=\langle \omega^2\rangle/2$. The mean lifetime of flow structures can be measured by the vorticity integral time scale, defined as: $T_\omega=\int_0^\infty C_\omega(\tau)\,d\tau$. We find that the normalized mean lifetime is more than doubled in the 20 ppm case compared to the buffer [Fig. \hyperref[fig3]{3(d)}, inset]. Overall, these results indicate that polymer stresses lead to larger structure size [Fig. \hyperref[fig3]{3(b)}] and longer flow memory [Fig. \hyperref[fig3]{3(d)}] in 2D bacterial turbulence.

\begin{figure}[t!]\label{fig4}
\centering
\includegraphics[width=2.6in]{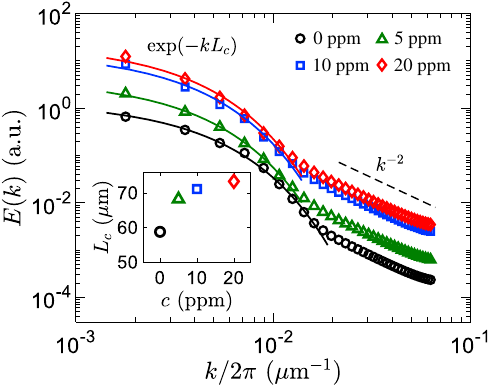}
\caption{Energy spectra $E(k)$ for different polymer concentrations. Solid curves are exponential fits to low-$k$ values, $\exp(-kL_c)$. A power law of $k^{-2}$ (dashed line) is drawn only for comparison. Inset: characteristic length scale $L_c$ obtained from the exponential fits, versus polymer concentration.}
\end{figure}

So far we have shown that minute amounts of polymers, while not imparting significant changes to the bulk fluid properties, are able to modify collective motion of swarming bacteria. One possibility is that the polymer molecules are locally oriented and stretched by the rotating flagella of the bacteria \cite{Patteson2015}. This can be estimated by calculating a (local) Weissenberg number, $W\!i=\lambda\dot\gamma$, which characterizes the degree of anisotropy in the shear deformation experienced by polymer molecules. Here, $\lambda$ is the fluid relaxation time and $\dot\gamma$ is the fluid local shear rate. Since the polymeric solutions are well below their overlap concentration ($c/c^* \ll 1$), we compute the Zimm relaxation time for dilute polymeric solutions \cite{Zimm_1956}: $\lambda_Z=FM_w\eta_{s}[\eta]/RT$, where $F$ is a factor \cite{Supp_Mat}, $M_w$ is the polymer molecular weight, $\eta_s$ is the solvent viscosity, $[\eta]$ is the polymer intrinsic viscosity, and $R$, $T$ are the gas constant and the temperature, respectively. We find a Zimm relaxation time of $\lambda_Z \approx 2$~ms \cite{Supp_Mat}. While this is a relatively short time scale, \textit{S. marcescens} possess rapidly rotating flagella generating significant fluid velocity gradients. A rough estimation \cite{Supp_Mat} yields a local Weissenberg number in the range of $W\!i\sim 2\,\textrm{--}\,20$, suggesting significant stretching of the polymer molecules near individual bacterium, and strong viscoelastic effects. This is supported by the experiments performed with tethered \textit{Escherichia coli}, which can stretch DNA molecules in the vicinity of their flagella \cite{Patteson2015}. The stretching of polymers may then mediate bacterial local interaction and lead to long-range hydrodynamic effects, which could explain the increase in structure size [Fig. \hyperref[fig1]{1} and Fig. \hyperref[fig3]{3(b)}].

To gain further structural insights into the bacterial turbulence, we examine kinetic energy distribution over different length scales by the energy spectrum, $E(k)=2\pi k\langle|\hat{\mathbf{u}}(\mathbf{k})|^2\rangle$, where $\hat{\mathbf{u}}(\mathbf{k})$ denotes the Fourier transform of the velocity field, $\mathbf{k}$ is the wavevector, and $k=\vert \mathbf{k}\vert$ is the wavenumber. While in inertial turbulence the energy spectra are routinely used to study the energy transfer across scales \cite{Frisch1995turbulence}, recent works demonstrate the absence of such energy cascade in active turbulence at low inertia \cite{Alert_NP_2020, Carenza_EPL_2020,Alert_Review_2022}. Instead, we view the energy spectra as a proxy for the spatial structure of swarming \textit{S. marcescens} suspensions and the associated velocity fields. As shown in Fig. \hyperref[fig4]{4}, at large scales (low $k$), $E(k)$ for all polymer concentrations are found to follow an exponential decay, $\exp(-kL_c)$. The characteristic length scale of the exponential decay, $L_c$, is of the same order as the integral length scale, $L_\omega$, and follows the same increasing trend with the polymer concentration (Fig. \hyperref[fig4]{4}, inset). Therefore, we identify the exponentially-decaying part of $E(k)$ with the large-scale collective motion of bacteria.

At smaller scales (high $k$), below the film thickness ($40~\umu$m), we report a power-law decay close to $E(k) \sim k^{-2}$ for all cases. The observed power law is inconsistent with the universal scaling laws recently reported for active matter in Newtonian fluids \cite{Giomi_PRX_2015,Alert_NP_2020, Carenza_EPL_2020,MartinezPrat_PRX_2021,Alert_Review_2022}. Within that scenario, the suspension is viewed as a ``gas'' of clusters: within a cluster, swimmers are nematically aligned with their neighbors, while at large scales, the directions of local nematic order are decorrelated. In the absence of confinement, the large-scale spectrum corresponds to that of a suspension of independent hydrodynamic dipoles \cite{Giomi_PRX_2015,Alert_NP_2020, Carenza_EPL_2020, MartinezPrat_PRX_2021,Bardfalvy_SM_2019,Morozov_PRX_2020}, $E(k)\sim k^{-1}$, while the small-scale spectrum below the cluster size is $E(k)\sim k^{-4}$ \cite{Giomi_PRX_2015,Alert_NP_2020,MartinezPrat_PRX_2021,Alert_Review_2022}. We, therefore, conclude that the active nematics spectra \cite{Giomi_PRX_2015,Alert_NP_2020,MartinezPrat_PRX_2021} and the underlying assumptions about the structure of the chaotic motion are inconsistent with our observations in swarming \textit{S. marcescens}. This discrepancy cannot be attributed to the rheological properties of the polymer solutions either, since simulations of dipole pusher swimmers in Oldroyd-B type viscoelastic fluids have found power laws of approximately $k^{-4/3}$ and $k^{-4}$ for low and high $k$, respectively \cite{Li_PRL_2016}, in line with the active nematics spectra. We note, however, that the scale at which the exponential decay gives way to the power-law decay in Fig. \hyperref[fig4]{4} is broadly consistent with the thickness of the film ($40$ $\umu$m), thus lending support to its hydrodynamic origins.

Finally, Fig. \hyperref[fig4]{4} also shows that the addition of polymers leads to an increase in spectral power at all wavenumbers due to the enhancement in bacterial swimming speed; the increase is, however, non-uniform due to the accompanied increase in flow structure size. These observations contradict recent theoretical work on collective motion in model viscoelastic fluids in 2D suspensions of motile organisms \cite{underhill_PRE_2011,Underhill_JNNFM_2014,Li_PRL_2016}, which predicted the reduction of the flow structure size and suppression of the energy content at large scales in the presence of polymers. Overall, our energy spectra show an exponential decay at large scale that is associated with the length scale of collective motion, and a power law of approximately $k^{-2}$ at smaller scales, with accompanied increases in spectra power due to larger swimming speeds.

In summary, we experimentally show that even minute amounts ($\le20$ ppm) of polymer additives can significantly alter the structure and dynamics of bacterial collective motion in dense active suspensions. As the polymer concentration is increased, we find an increase in the average swimming speed but velocity fluctuations are systematically suppressed (Fig. \hyperref[fig2]{2}). These are not merely flow modifications; the presence of polymers in the bacterial swarms leads to different flows. One piece of evidence is that the velocity distributions of the swarms in the polymeric solutions are approximately Gaussian, while these in the buffer case (without polymer) are super-Gaussian. Surprisingly, the addition of polymers leads to an increase in the size and lifetime of coherent flow structures. This is in contrast to the trends found in numerical simulations \cite{Li_PRL_2016, Underhill_JNNFM_2014}, where polymers decrease the size of flow structures. Overall, our work provides insights into the collective dynamics of microswimmers in complex fluid environments, in particular those containing polymers such as mucus and biofilms.

\vspace{+1ex}

We thank Daniel Blair and Christopher Browne for helpful discussions. This work was supported by NSF Grant No. DMR-1709763.


\bibliographystyle{apsrev4-2}
\bibliography{reference.bib} 
\end{document}